\documentclass[11pt,twoside]{article}
\usepackage{asp2010}

\resetcounters

\begin{document}

\title{NMAGIC made-to-measure particle models of galaxies}
\author{Lucia~Morganti \& Ortwin~Gerhard
\affil{Max-Planck-Institut f\"ur Extraterrestrische Physik,
Giessenbachstra{\ss}e, D-85741 Garching, Germany}}

\begin{abstract}
The parallel code NMAGIC is an implementation of a particle-based method
to create made-to-measure models in agreement with observations of galaxies.
It works by slowly correcting the particle weights of an evolving N-body system,
until a satisfactory compromise is achieved 
between the goodness of the fit to a given set of observational data, 
and some degree of smoothness (regularization) of the underlying particle model.

We briefly describe the method together with
a new regularization scheme in phase-space,
which improves recovering the correct orbit structure in the models.
We also mention some practical applications
showing the power of the technique in investigating the dynamics of galaxies.

\end{abstract}

\section{Introduction}
In the last decades, an increasing amount of high quality photometric and kinematic 
data for galaxies have become available.
The dynamical state of the observed galaxy, however, cannot be directly
inferred from observations, due to projection effects,
and modelling is essential to learn about the distribution of stellar orbits and the total
(i.e. due to luminous and dark matter) gravitational potential in the observed galaxy.
Therefore, several methods to model the observational data 
and create "made-to-measure" models
have been devised.

For instance, assuming that  all the integrals of motion are known,
one can fit observations with parametrized distribution functions
\citep{dejonghe84,dejonghe86,bishop87,gerhard91,hunterdezeeuw92,carollo95,kuijken95, mago95, merritt96,dehger93, matger99},
or solve the Jeans equations subject to the observational constraints
 \citep{binmam82,binney90,magobin94,lokas02,cappellari08}.
 
Another technique which is widely used is the 
Schwarzschild orbit superposition method \citep{sch79, sch93}:
a large library of orbits is computed in a fixed potential,
and then the weights of individual orbits are adjusted
until the model matches the photometry and kinematics of the target galaxy
\citep{richstone85,  
cappellari02, gebhardt03, 
Valluri04, thomas05, vandenbosch10}.

\citet{st96} proposed an alternative, particle-based method.
A modified version suitable for modelling observational data with errors 
was designed by  \citet{dl07} and implemented in the parallel code NMAGIC.
More recent implementations of this method can be found 
in \citet{dehnen09} and \citet{longmao10}.
The basic idea is to evolve a system of particles
while slowly correcting the individual weights of particles
until the $N$-body system reproduces the observational data.
The method is very powerful, since no orbit library needs to be
computed or stored, no symmetry restrictions need to be
imposed, and the potential can be evolved self-consistently from the particles.
Moreover, the algorithm properly accounts
for observational errors, and a great variety 
of observational data can be used simultaneously in the weight adaptation,
including photometry, long-slit spectroscopic data, 
integral field absorption-line kinematics, and PNe velocities.

So far, the particle made-to-measure method has been used to investigate 
the dynamics of the Milky Way's bulge and disk \citep{bissantz04},
and the dark matter fraction and orbital structure 
in the outer halos of elliptical galaxies \citep{dl08,dl09,das11}.

\section{Made-to-measure particle models of observational data}
An $N$-body system of particles is trained 
to reproduce the observables of a target galaxy
by maximizing the merit function
\begin{equation}
F = -\frac{1}{2}\chi^2+\mu S
\label{eqn:SF}
\end{equation}
with respect to the particle weights $w_i$
\citep[see e.g.][]{st96,dl07}.
Such maximization strikes a compromise
between the goodness of the fit $(\chi^2)$ in terms of deviations
between target and particle model observables,
and a pseudo-entropy
\begin{equation} 
S = -\sum_i w_i \log (w_i/\hat{w}_i),
\label{eqn:entropy}
\end{equation}
which serves the purpose of regularization.
Since typically the number of particles is much higher 
than the number of noisy and sparse constraints on the particle model,
the problem is intrinsically ill-conditioned,
and some regularization is necessary.
This is achieved by pushing particle weights towards 
a smooth distribution of  predetermined priors $\hat{w}_i$. 

The maximization of equation~(\ref{eqn:SF}) 
translates to a prescription for adapting the weights
of the particles while they are evolved in the gravitational potential,
which can be either fixed and known a priori, 
or time-varying and self-consistently computed.

\subsection{Regularizing particle models}
Traditionally, priors are set to $\hat{w}_i=w_0=1/N$
(the ``flat'' priors in Bayesian statistics), 
and are kept constant during the modelling.
Likewise, the individual weights of the initial particle model
are usually set to $1/N$.

However, as discussed in \citet{mg12}, 
smoothing the weights globally 
towards a set of preassigned, constant priors 
makes it difficult to construct made-to-measure galaxy models
with both smooth and anisotropic DFs 
unless the initial particle model is already close
to the target galaxy, i.e.~the dynamics of the galaxy is already known.
In this paper, we described and implemented in NMAGIC a new regularization method
based on moving (i.e.~not constant) priors
which follow the smooth phase-space structures 
traced by the weight distribution, as the latter evolves to match the observational data.
The computation of such new individual priors 
is particularly easy when the integrals of motion are known, 
or at least one can compute approximately conserved quantities from the particle orbits.
In \citet{mg12} we consider spherical and axisymmetric cases, 
and show that the procedure facilitates recovering 
both a smoother and more accurate DF 
without erasing global phase-space gradients.

\section{Convergence to a theoretically unique solution}
It is a natural question whether the made-to-measure modelling technique 
can recover the target properties from observations
independently of the initial particle model.
As we show in \citet{mg12}, if a unique inversion of data 
to recover the underlying target DF exists \citep{dejonghe92},
then it can actually be found via NMAGIC modelling.
In Fig.~\ref{fig1} the quality to which the orbital anisotropy
of the target galaxy can be recovered in such a case
for different initial particle models can be appreciated.
\articlefigure[angle=-90.0,width=0.98\hsize]{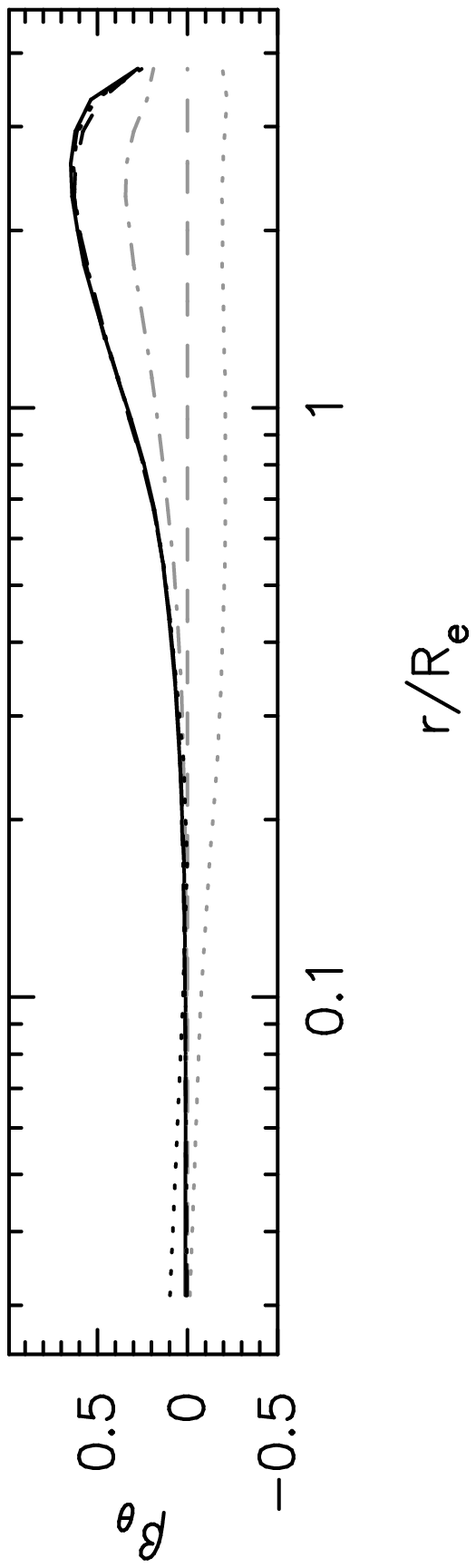}{fig1}{Recovery of the anisotropy parameter of a spherical target galaxy (full black line) 
from idealized line-of-sight velocity data. Radius is normalized by
the model's half-light radius $R_{\rm e}$. 
The black lines, nearly on top of each other, show the final NMAGIC models obtained with 
our new regularization scheme using different initial particle models 
(corresponding grey lines).}

By contrast, lack or poor quality of data
introduce degeneracies in the dynamical modelling results;
see \citet{mg12} for a quantitative analysis.

\section{Modelling the halos of elliptical galaxies}
The outer halos of elliptical galaxies are particularly interesting because
they are dark matter-dominated,
and because the orbital structure of stars there
more strongly preserves the imprint of formation mechanisms,
due to longer dynamical time scales.

Dynamical models of these galaxies were constructed with NMAGIC, 
fitting a wide variety of photometric and kinematic data
in a sequence of gravitational potentials for the dark matter halos.
When modelling the intermediate luminosity ellipticals
NGC~4697 and NGC~3379, \citet{dl08,dl09} found
that a variety of potentials are consistent with the data,
showing that the well-known mass-anisotropy degeneracy
\citep{binmam82} remains substantial when the velocity dispersion
strongly decreases with radius, even with data out to more than $5  R_{\rm e}$.
By contrast, the mass distribution in luminous ellipticals
with extended flat dispersion profiles is well-constrained \citep{das11}.

\section{Conclusions}
The parallel code NMAGIC is a powerful tool to build
made-to-measure galaxy models, which reproduce
a wide variety of observational data.
NMAGIC works by adapting the weights of an $N$-body particle system 
until the target observables are well matched, subject to additional regularization constraints.

We have briefly described a new, improved regularization scheme in phase-space,
and shown how well the intrinsic properties of the target galaxy can be recovered
independently of initial conditions,
and how NMAGIC models can help us
in learning about the orbital structure and gravitational potentials in galaxies.
\bibliography{Morganti_L}

\end{document}